\def\cena{Centaurus~}
\def\scul{Sculptor~}
\def\etal{{\it et al. }~}
\def\Msun{\mbox{M$_\odot$}~}

\def\Msunpc2{\mbox{M$_{\odot}$~pc$^{-2}$}}
\def\Msunpc3{\mbox{M$_{\odot}$~pc$^{-3}$}}
\def\Msolpc2{\mbox{{\rm M}$_{\odot}~{\rm pc}^{-2}$}}

\def\Lsunpc2{\mbox{$L_\odot$~pc$^{-2}$}}

\def\ie{{\it i.e.~}}

\def\scg{Sculptor and Centaurus groups~}

\def\kms{\mbox{\,km\,s$^{-1}$}}

\def\arcsec{\mbox{$^{\,\prime\prime}$}}
\def\arcmin{\mbox{$^{\,\prime}$}}
\def\degrees{\mbox{$^{\circ}$}}

\def\angst{\mbox{$\,${\rm \AA}}~}
\font\twelverm=cmr12
\font\ninerm=cmr9
\catcode`\@=11
\def\gapprox{\mathrel{\mathpalette\@versim>}}
\def\lapprox{\mathrel{\mathpalette\@versim<}}
\def\@versim#1#2{\lower2.45pt\vbox{\baselineskip0pt\lineskip0.9pt
    \ialign{$\m@th#1\hfil##\hfil$\crcr#2\crcr\sim\crcr}}}
\catcode`\@=12

\def\sun{\ifmmode\odot\else$\odot$\fi}
\def\up#1{\leavevmode\raise0.16ex\hbox{#1}}
\def\[{\up[}
\def\]{\up]}

\def\OIIIf{\hbox{\twelverm \[O\thinspace\ninerm III\twelverm\]}~}
\def\Halpha{\hbox{\rm H$\alpha$}}
\documentstyle[12pt,aaspp4]{article}

\slugcomment{Submitted to AJ.}

\lefthead{C\^ot\'e \etal} {}
\righthead{Dwarf Galaxies in Nearby Groups}

\begin{document}

\title{Discovery of Numerous Dwarf Galaxies \\
    in the Two Nearest Groups of Galaxies}

\author{St\'ephanie C\^ot\'e}
\affil{Mount Stromlo Observatory, P.M.B. Weston Creek, ACT2611, Australia, and 
 European Southern Observatory, Karl-Schwarzschild Str. 2,
    D-85748 Garching, Germany}

\author{Kenneth C. Freeman}
\affil{Mount Stromlo Observatory, P.M.B. Weston Creek, ACT2611, Australia} 

\author{Claude Carignan}
\affil{D\'epartement de Physique et Observatoire du Mont M\'egantic, Universit\'e de 
   Montr\'eal, C.P. 6128, Succ. A, Montr\'eal, Qu\'ebec H3C 3J7}

\author{Peter J. Quinn}
\affil{Mount Stromlo Observatory, P.M.B. Weston Creek, ACT2611, Australia, and 
 European Southern Observatory, Karl-Schwarzschild Str. 2,
    D-85748 Garching, Germany}

\begin{abstract}

We report the discovery of a large population of dwarf irregular galaxies in the
two nearest groups of galaxies outside the Local Group, the Sculptor and
the Centaurus A groups (2.5 and 3.5 Mpc).  Total areas of approximately 940 and 910 square degrees 
in these groups
were scanned visually on the SRC J films to find dwarf candidates. Redshifts 
were obtained by an 
HI survey carried out at Parkes, with detection limits of     
$4 \times 10^6$ \Msun and $7.8 \times 10^6$ \Msun , and
33 dwarf galaxies were successfully detected in the groups.
A follow-up optical survey (\Halpha ~spectroscopy) detected a few more, and 
confirmed most of the HI 
redshifts. A total of 16 and 20 dwarf galaxies are found in Sculptor and 
Centaurus A, of which 
 6 are newly identified objects, and 5 more have newly determined redshifts.

In both groups the dwarf members show a wider spatial and velocity
distribution than the brighter members. From their radial velocities and
projected distances we estimate the crossing times of the groups, which
confirm that they are not yet virialised.

\end{abstract}

\clearpage
\section{Introduction}

Dwarf galaxies are by far the most numerous type of galaxies, not only
in the Local Group but also in the nearby clusters which have been studied in
detail (eg. Sandage \& Binggeli 1984 for Virgo; Caldwell \& Bothun 1987
for Fornax). It should therefore be normal to assume that nearby groups of
galaxies contain similarly a large population of dwarfs.
However it is often believed that these nearby groups are devoided of
gas-rich dwarfs at least, based on earlier HI surveys of Lo \& Sargent
(1979) and Haynes \& Roberts (1979).  Yet these surveys were done
with only very sparse samplings in the case of the Sculptor group, and 
to sensitivities of a few times 10$^7$ M$_\odot $ for the M81, CVnI,
and NGC 1023 groups. Finding nearby dwarf galaxies is crucial in
defining the faint-end of the luminosity function of galaxies, which
has far-reaching consequences in the study of faint galaxy populations
and galaxy formation in general. But redshift surveys of
apparent-magnitude-limited samples are, for various reasons, not very 
 efficient at finding low-luminosity systems. 

 The \scul group and
the \cena A group of galaxies are the most nearby groups 
 in the southern Hemisphere. Being roughly 3 magnitudes closer than the
 Virgo and Fornax clusters they bring us a unique opportunity to study
 the very faint end of the luminosity function. The goal of this project
 was therefore to investigate the dwarf population in these two groups,
 especially the gas-rich dwarf irregulars (dIrrs). It is much
 easier to get redshifts for dIrrs (in HI) than to determine redshifts
 for dEs candidates, and also our main interest was 
 to carry out dark matter studies on dIrrs.

The \scul and \cena A groups offer very different environments to their
dwarf population. The \scul group is mainly composed of 
late-type spiral galaxies. Its five major members, NGC 55, NGC 247, NGC
253, NGC 300, and NGC 7793, are almost all normal gas-rich systems, and their
properties are listed in Table~\ref{t:sculm}. Our adopted distance,
based on numerous distance indicators (listed in Puche \& Carignan 1988)
will be 2.5 Mpc. Two more galaxies, NGC 45 and NGC 24, also late-type
spirals, are found in the same region but several Mpcs further away,
possibly forming an extension of the group.
 Amongst the Sculptor members only NGC 253 shows
starburst activities while the other galaxies are quiescent. Their HI
distributions have been mapped in details at the VLA by 
 Carignan \& Puche (1990a, 1990b), Puche \etal (1990, 1991a, 1991b).
 Miller (1996) has imaged in H$\alpha $ the known dwarf
 members of the group and detected only two with current star
 formation.
The \cena A group on the other hand is a loose chain of galaxies, a
heterogeneous assembly of early to late-type galaxies, having the largest
dispersion of morphological types amongst all of the 55 nearest groups
(de Vaucouleurs 1979).
  Almost all of the major members are
 abnormal. Not only is NGC 5128, Centaurus A itself, one of the most
 peculiar radio galaxies, but the other main members also show signs of
 activities. For example NGC 5236 (M83) has a starburst nucleus and
 possesses an anomalously high supernovae rate (Telesco \& Harper 1980),
 and an asymmetric HI velocity field (Lewis 1969). NGC 5253 also
 shows starburst activities and a high supernovae rate (eg. van den
 Bergh 1980), and 
  a very disturbed H$\alpha$ velocity field (Taylor 1992).
NGC 5102 is an unusual S0 galaxy in a post-starburst phase (Pritchet
1979). And NGC 4945 has a Seyfert nucleus emitting variable hard X-rays
(Moorwood \& Olivia 1994).
  Several authors have suggested that these
peculiarities are induced by accretion of gas-rich dwarf systems (Graham
1979; van Gorkom \etal 1990). Therefore the study of the gas-rich dwarfs
in this group is of particular interest.
 The main members are listed in
Table~\ref{t:cenam}, and the adopted distance for this group is 3.5 Mpc 
based on the newly derived distances for these galaxies (see the Table). 
It is therefore
the second nearest group outside the Local Group, more or less ex-aequo
with the M~81 group (from Cepheids M~81 is placed at 3.6 Mpc, Freedman \etal
1994).

This paper, the first of a series, presents the redshift
surveys for finding dIrrs in Sculptor and Centaurus A. Further papers will 
present full B,R,I photometry of
these nearby dwarfs (C\^ot\'e \etal 1997a), the neutral hydrogen kinematics of 
a sample of these objects (C\^ot\'e \etal 1997b) and finally the dark matter 
studies for that sample of dIrrs (C\^ot\'e \etal 1997c).

\section{The Survey}
\subsection{Visual survey}

The first step was to find dwarf galaxy candidates in the
region of the Sculptor and Centaurus A groups from visual inspection of 
SRC J survey films.
 These groups,
 because of their proximity, subtend very large angles on the sky.
The areas on which to select candidates 
 were determined from the sky coverage of the already-known galaxy members
in each group (main members as well as dwarf members known at the time).
 The fields thus 
 retained for visual scanning were those with centers between:
\begin{equation}
23^h \leq \alpha  \leq 2^h \;;  
\; -20\degrees \geq \delta \geq -45\degrees
\end{equation}
\begin{equation}
12^h30 \leq \alpha  \leq 15^h \;;  
\; -20\degrees \geq \delta \geq -50\degrees
\end{equation}
for the \scul and \cena A groups respectively, where $\alpha $ and $\delta $ are the right ascension and 
 declination at epoch 1950. This corresponds to wide areas of approximately  
 50 SRC survey films in each case.

Each film was inspected visually with a small magnifier, scanning once North
to South, then West to East, to locate all objects which could be  
irregular galaxies at low redshift. 
 A dwarf irregular should be easily 
identifiable at such a small distance of a few Mpc, showing much  
structural details, with even the brightest stars resolved.
We selected particularly all 
low-surface-brightness galaxies which showed some degree of stellar
resolution, or some distinct \ion{H}{2} regions or condensations, including 
all objects 
 with small bulges or some form of high-surface-brightness central regions, 
and those composed of bright knots embedded in low-surface-brightness
envelopes.
 The objects excluded from the candidate list were giant ellipticals and
obvious distant giant spirals (judged as such from the relative sizes of the bulges
and the tightness of the spiral arms).
 With these criteria 
we should have included in our candidate lists all gas-rich dwarfs like dwarf spirals, dwarf
irregulars, and blue compact dwarfs (BCD) as well, although some BCDs 
that appear as a
single symmetric burnt-out clump on the SRC films with no associated 
low-surface-brightness 
 envelope would have escaped us, being indistinguishable from
background ellipticals.

A total of 123 dwarf galaxy candidates for the Sculptor Group and 145 for the
Centaurus A Group were thus identified, with no redshifts  
 in existing galaxy catalogues at the time, including 
 the RC2 (de Vaucouleurs \etal 1976), the {\it Nearby Galaxies Catalog} (Tully 1988), 
the {\it ESO Catalog} (Lauberts 1982), the {\it Southern Galaxy Catalog} 
 (Corwin \etal 1985), the {\it HI Catalog of Galaxies} 
(Huchtmeier and Richter 1989), and the NASA/IPAC Extragalactic 
Database\footnote{The NASA/IPAC Extragalactic Database (NED) is operated
by the Jet Propulsion Laboratory, California Institute of Technology, under
contract with the National Aeronautics and Space Administration.} (NED). 
These candidates are spread on the sky over approximatly 940 and 910
square degrees.
Amongst these, 58 objects are identified for the first
time and are named with the sigla 'SC' or 'CEN'  (see Tables~\ref{t:sback} and
~\ref{t:cback}). 

\subsection{HI survey}

Dwarf irregular galaxies being normally gas-rich, the easiest way to determine
their redshift 
 is with 21cm HI line observations. Some of these galaxies like DDO 154 for 
 example (Carignan \& Freeman 
1988) have actually 5 times more mass in HI than in stars. And at such low
redshift it is not very time-consuming to get down to an interestingly low
detection limit (in terms of HI solar mass) using a single-dish telescope. 
Also since the groups are so widespread on the sky there is little
chance of confusion even with a large beam.
The HI observations were carried out in August 1990 for the \scul candidates 
and February 1991 for the \cena A ones, at the 64m Parkes 
Radiotelescope. The telescope had a 
half-power beamwidth of 14.8\arcmin ~at 21cm and a sensitivity of 0.63 K/Jy.
The front end was a cryogenic FET receiver yielding a system temperature of
$T_{sys} \sim $40 K. The back end was the 1024 channel digital autocorrelator
(Ables \etal ~1975), configured to give two 512 channels spectra  (1 spectrum
for each of two orthogonal polarizations), with a bandwidth of 10 MHz.
This resulted in a channel separation of 4.1 \kms ~and a velocity resolution
after Hanning smoothing of 8.2 \kms , covering a velocity range of roughly
$-$200 to 1800 \kms . This setup ensured that we were surveying a wide enough
velocity interval for these two groups whose main members have heliocentric
velocities between 116 and 669 \kms , as well as preserving a good enough 
resolution to distinguish dwarf galaxies profiles. These do not necessarily
have large rotation velocities and therefore their profiles can often look
 like asymmetric gaussians, of much smaller widths than for spirals 
 (see Karachentseva 1990 who classified 539 dwarf HI profiles).
  The flux density calibration was carried  out by observing Hydra A 
(PKS 0915-18), with an adopted flux density of 43.5 Jy at 
21cm, and the velocity scale was checked on the source UKS 1908-621 which has
a radial heliocentric velocity $V_{sys} = $946 \kms ~and a profile width of
$\Delta V_{20} =$ 92 \kms .
Each observation consisted of a 10 minutes integration on source, alternated
with a 10 minutes integration on a sky reference position lying 10 minutes
away to the east, so that the same position (in hour angle) was observed during this  
reference scan to get the same background.

The reductions were performed with {\it SLAP}, an interactive spectral line
reduction package developed at Jodrell Bank, and {\it POSP}, developed at
NRAO. The sky 
reference spectra were used to subtract the sky and remove large baseline
variations, and the two polarization spectra were then co-added. 
 This baseline was interactively determined by masking out
any emission and fitting a low order chebyshev polynomial. The zeroth and
first moments of the emission were calculated to obtain the total flux and
heliocentric velocity of the galaxy. Line widths were determined at the 
20\% and 50\% peak flux levels. The resulting spectra had  rms values
typically around 30 mJy, 
 and so our 3$\sigma $ detection limits in \scul and \cena A are $4 \times 
10^6$ \Msun and $7.8 \times 10^6$ \Msun (for our adopted distances of 2.5 Mpc
and 3.5 Mpc respectively). Tables~\ref{t:sback} and ~\ref{t:cback} list
for all the HI detections the heliocentric velocity of the galaxy, the
HI integrated flux, and the linewidth at the 20\% level. The velocity
quoted is an average of the intensity-weighted velocity, and the midpoint
velocities at the 50\% and 20\% levels. Errors on these
velocities are typically 2 \kms . 
Figures~\ref{f:sallsp} and ~\ref{f:callsp} 
show the reduced Parkes spectra of all the detected objects over the
whole velocity range.

However one needs to be cautious when detecting HI emission in the velocity range 100 \kms 
~to 600 \kms ~in the direction of the \scg :
many High Velocity Clouds (HVC) associated with our Galaxy are known to have
 velocities sometimes as high as 400 \kms ~(Wakker 1990). So it could happen that such a HVC 
is lying in the 15\arcmin ~Parkes beam towards a 
candidate object (which could actually be a background object), confusing us into believing that the HI detected belongs to that galaxy. In fact there
are some HVCs with high positive velocities known to lie in the ({\it l,b}) area of 
both groups (Wayte 1990; Wakker 1990). These are not the only possible 
source of confusion: intra-group HI clouds have been detected in the \scul  group
by Haynes 
and Roberts (1979) who made extensive mapping in that region with
the Green Bank 140ft and found several seemingly `free-floating' HI clouds.
 From their observations it is not clear actually if these
clouds are genuine intergalactic HI clouds, mere HVCs, or some component of the 
Magellanic Stream (Mathewson \etal 1974), but for our 
purposes all these possibilities are to be considered HI pollution.
Therefore in order to discriminate genuine nearby dwarf HI emission from HI HVCs or
clouds, further 5-point mappings were carried out in May 1991 and 
September 1991. This consists of integrating 
15\arcmin ~away (one beam width) from the source to the North, South, East
and West (or in a cross pattern: North-West, North-East, etc). In the case of a dwarf 
 galaxy the HI emission will be mostly concentrated
in the middle beam, and velocity variations between different pointings will 
indicate rotation; for a HI cloud the extended emission is less ordered
in velocity and on the
sky. It is also very useful for the dwarfs that lie near a HI-rich
large galaxy like M83 or NGC 4945, where one can detect HI flux in the
sidelobes of the beam and so one component in the spectrum will be due
to the HI envelope of the large galaxy; in this case a 5-point mapping helps 
confirming that there is an independent object as well in the spectrum.
 However this 5-points method works only if there is actually
some emission 15\arcmin ~away and so is not always a useful discriminant  
between dwarfs and usurper HI clouds.
The clearest way to discriminate between HI pollution and genuine dwarfs is 
to try to get a redshift optically in H$\alpha $ for these objects. The 
presence of H$\alpha $ at the same velocity as HI indicates the presence of 
star formation, therefore ruling out a neutral cloud.

\subsection{H$_\alpha $ survey}

The H$\alpha $ spectroscopy observations were carried out in April
1991, August 1991 and November 1991, with the Double Beam Spectrograph (DBS)
on the MSSSO 2.3m telescope. Not only did we observe the objects detected in HI
at the groups' velocities (ie: the objects that needed confirming because of 
HVCs etc), but also all the objects on our catalog lists which 
were not detected at all in HI, in the hope to identify new dwarf galaxies
weak in HI but detectable in H$\alpha $. The long slit was about 6\arcmin ~and
was set to a width of 3\arcsec , positioned when possible along the 
major axis of the object. The detector used was a photon--counting--array
with a spatial resolution of 0.67\arcsec ~per pixel. The 1200 G/mm 
gratings were used in the blue and red arms, covering spectral ranges of
4950 to 5250 \angst in the blue, and 6450 to 6750 \angst in the red, at 0.4 
\angst per pixel (18 \kms ~at H$\alpha $).  This blue arm setup was to target 
the \OIIIf  4959 and 5007\angst lines, but 
 these lines being always fainter than the H$\alpha$, 
 in most cases the velocities were determined from the H$\alpha $ alone.
Exposures were 2000s per object, or were stopped before that when sufficient
signal-to-noise was achieved for a redshift determination.

Spectral reduction was carried out with the FIGARO package from K. Shortridge 
(AAO). Each galaxy observation was bracketed with arc lamp exposures for
wavelength calibration. 
 Sky spectra were extracted from the frames at regions well outside
galactic emission. Gaussian profiles were then fit to the emission lines, and a
heliocentric correction was applied to the resulting velocities.
These are listed in Tables~\ref{t:sback} to ~\ref{t:cfin}, and have errors
of typically 10 \kms.

\section{Results}
\subsection{Membership of Sculptor and Centaurus A}

Of the 123 and 145 original candidates in Sculptor and Centaurus A, all observed, 
we obtained new redshifts for 108 objects. For 61 other objects,     
redshifts were found in the literature during the course of the project,  
 for example when the RC3 became available, and for simplicity are not 
repeated in our Tables~\ref{t:sback} and ~\ref{t:cback}; they can be found 
in the RC3, Maia \etal (1993), Dressler (1991), Da Costa \etal (1991), Fouqu\'e
(1990), Rhee (1992), and Parker (1990). 
  Finally 74 objects remain 
 redshiftless.

A total of 16 dwarfs are found to be members of the Sculptor group and 20 of
the Centaurus A group, and are listed in Tables~\ref{t:sfin} and ~\ref{t:cfin}. 
Amongst these, 25 were already known in the literature
(we confirmed their redshift), 5 more were already identified and catalogued 
but without known redshifts, and finally 6 of these objects are identified for
 the first time. None of these 6 new objects had been detected by 
 IRAS. These 36 members of Sculptor and Centaurus A include all our
objects with confirmed redshifts in H$\alpha $, as well as those detected only
 in HI but with successful 5-points mappings. Only 21 of these dwarfs
had their HI emission confirmed with an H$\alpha $ redshift. And 3 more
objects, not detected in HI, were found in the H$\alpha $ survey:  
 AM 0106-382 in \scul , as well as NGC~5206 and ESO 272-G025 in \cena A.  
Further deeper integrations in HI at Parkes still did not detect any neutral
gas content, bringing their HI upper limit to 
 $M_{HI} < 2.8 \times 10^6 ~\Msun$ for 
 \scul and $M_{HI} < 5.5 \times 10^6 ~\Msun $ for \cena A.
Several HVC's were indeed found in the line-of-sight of some of our 
candidates as suspected: 
 in all, 15
HI detections were rejected from our dwarf lists, either after the 5-point
mapping, or after obtaining an H$\alpha $ detection at higher redshift.
These HVC HI profiles are considerably more messy that the confirmed galaxies'
profiles. We are therefore confident that the 12 dwarfs which remain 
unconfirmed in H$\alpha $ but were kept in the sample judging from their 
5-point mapping are bona fide dwarfs. Moreover, subsequent surface photometry, 
 to be presented in C\^ot\'e \etal (1997a), shows
that these objects have the same blue mean colours as our confirmed
dwarfs sample.

Amongst other known objects in the literature to be found in the Sculptor
region is UGCA 438  
 (alias UKS 2323-326), which we decided not to include as a Sculptor member: 
 Longmore \etal (1978) who
discovered it estimated a distance of 1.3 Mpc using the brightest stars, 
although van den Bergh (1994) 
claims it does not belong to the Local Group, based on its location on the 
(heliocentric velocity) versus (distance from the solar apex) diagram. So it 
might
be lying in fact  
 between the two groups. Two other controversial cases are   
 ESO 294-G010 and ESO 410-G005 which were included in Miller's (1994) work
on Sculptor dwarfs, but which we have not considered  members 
here, because ESO 294-G010 shows a 2.5$\sigma $ detection at 4450 \kms ~and 
ESO 410-G005 was not detected neither in HI nor in H$\alpha $ (we have prefered 
to include only the objects proven to be in the groups rather than those that 
have not been proven to be in the background).

It should be stressed again that only positions of candidates selected
visually on the films were observed, ie: no Parkes HI integrations at random
positions on the sky were performed in the regions of the two groups.
It is therefore more than probable that many more faint dwarfs are to be
found in these regions. A survey with the Parkes 21cm Multibeam receiver
for example could provide a more complete sample down to an interesting HI mass
limit with full coverage over the area of the groups.

The spatial and velocity distribution of the 36 confirmed dwarfs of Sculptor
and Centaurus A are shown in Figures~\ref{f:dis} and ~\ref{f:vlg}, and will be 
discussed at greater length in section 3.3.

\subsection{Morphology of the dwarfs}

 All the confirmed dwarfs are shown in Figure~\ref{f:atlas} in a
montage of enlarged SRC J prints.
  Their optical 
sizes range from about 40\arcsec ~up to 11\arcmin ~(0.5 kpc to 11 kpc).
This atlas reveals a large diversity of 
 morphologies, although a few subclasses of dwarfs are noticeable. 
 ESO 383-G087  is certainly a dwarf spiral   
(type Sm), being in fact one of the rare dwarf galaxies with clear spiral 
structure. Other dwarf spirals are ESO 274-G001, which even has a small
nucleus, while DDO 6, DDO 161, and DDO 226 show hints of a bar, 
but no nuclei. A few objects can be classified as BCDs (like  
  ESO 347-G017, NGC 5264 or NGC 5408), harbouring compact regions of very 
high surface brightness, although only ESO
347-G017 satisfies 
the strict Thuan and Martin (1981) criterion
that the localized star-forming region should be $<$ 1 kpc.
ESO 272-G025 and AM0106-382 are possibly extreme BCDs: they have the most compact
star-forming regions, and are amongst the only three objects which were not 
detected in HI, perhaps because the gas has been succesfully expelled as is
often believed to happen in low-mass BCDs during their burst of star formation.
Most of our objects are gas-rich dwarf irregulars, with sparse \ion{H}{2} regions 
either concentrated in the center of the low-surface-brightness envelope (ESO
324-G024), or dispersed through the whole envelope (ESO 381-G020).
Amongst these LSB objects, some like DDO6 or ESO 293-G035 have an almost  
`cometary' look, with 
a clump of \ion{H}{2} regions at one extreme of a otherwise LSB disk. 
 These particular `cometary dwarfs' are actually very 
similar to many faint galaxies now revealed in the HST Deep Field.
BCDs with similar morphologies were also found by Loose \& Thuan (1985),
who suggested that it is due to self-propagating star formation
 which stopped at the edge of the galaxy.

 These cometary dwarfs contrast with dIrr objects like ESO 444-G084 or SGC142448-4604.8
 that are 
of extremely low surface-brightness overall.
 Some objects like SC2 or CEN6 are just too `dwarfish' to 
discern any features but appear significantly distorted.
 Finally, although this survey was focusing
on gas-rich dwarfs, some early-types were caught in the sample:
 NGC~5206, not detected in HI, is a rather symmetrical-looking dE, whose light 
profile is well-fitted by a de Vaucouleurs $r^{1/4}$ law (Prugniel \etal 1993). 
Following Binggeli and Cameron's (1991) classification for early-type dwarfs,   
NGC 59 qualifies to be a dS0 type B (for its high flattening) and NGC 5237 a dS0
type D (for its boxiness).
Surface photometry results
will be presented in C\^ot\'e \etal (1997a), which will also present 
the luminosity function obtained, and discuss the selection and completeness 
of the survey.

\subsection{Spatial and velocity distribution}

The distribution on the sky of the member dwarfs is shown for each group
in Figure~\ref{f:dis}. In both groups the dwarfs are more widely spread  
than the brighter main galaxies. 
 Despite the difference of environment in the two groups 
 the spatial coverage of the dwarfs is found to be rather similar for
both. In each case an appendage of dwarfs is spearing away from the main 
members, while a few other objects have gathered near massive members, 
like NGC 247 or M83, and even though these dwarfs are not all bound to 
these bright galaxies they probably respond more to these galaxies' potential
then to the overall-group potential. Interestingly just as the
main \cena A members are aligned in a long chain, the dwarf members show also such
a elongated structure albeit heading more in a south-east direction. In the  
\scul group as well the dwarfs extend out to the south-east.  This is not
 due to some selection bias related to  
 obscuration effects,  
because for \scul the galaxy NGC~253 is very near the South Galactic Pole 
(at $b=-88$\degrees), while for \cena A 
the galactic plane lies to the south (NGC~4945 is at $b=13$\degrees). 
In fact the distribution of candidate objects from our original lists is
very uniform in the regions surveyed (for \scul ~it is even slightly
skewed towards the North).

Figure~\ref{f:vlg} shows the velocity distribution of the dwarfs in each
group, where their declination is plotted against their velocity $V_{LG}$
($V_{LG}$ is the velocity relative to the Local Group, which is more 
useful than the heliocentric velocity because of the groups being so
spread out on the sky: we used the Yahil \etal 1977 formula for the
correction). Here again one 
can see how the dwarfs have a wider velocity coverage than their bright
companions. However it is quite clear from this figure where each group 
cuts off in velocity-space, \ie : the dwarfs do not cover the whole observed
velocity range uniformly. Both groups are well `contained' in velocity space,
not overlapping with any string of stray dwarfs. Only at $V_{LG} > 1000$ \kms 
~appears another agglomeration of galaxies, which Tully's catalogue
(1988) refers to as the `Centaurus Spur'. This is similar to what is
observed in more extensive surveys, in which dwarfs are loosely
following the bright galaxies distribution, and are found on the edges
of the voids but do not seem to be filling these voids (see for example
Thuan \etal 1991).

The fact that the dwarfs distribution is wider spatially  
 than that of the more massive 
members is reminiscent of the situation in the Local Group, where
 most of the dIrr galaxies sit in regions of low density, out at the
fringes of the group. 
 A similar behaviour has been observed as well for the late-type
dwarfs of nearby clusters, like Virgo for which Bothun \etal (1985) conclude 
that the normal late-type spirals as well as the dIrrs form an extended cluster population 
that has not yet fallen
into the Virgo core (see also Binggeli \etal 1987).
 Also observed in the
Virgo cluster is a general HI deficiency in the spirals, 
  which also have smaller characteristic HI sizes than
 field spirals (Cayatte \etal 1994). Here for \scul and \cena A there are
 no significant changes in the HI properties of dwarfs at the 
 periphery of the group compared to those near the center of mass of
 the group or orbiting a bright member, and in such small and loose
 groups one does not expect to encounter the gas--removal processes that
 operate in denser groups or clusters.

\subsection{Group kinematics}
 
 The sample of radial velocities of the dwarfs and their projected distances
from the center of mass of each group can be used to estimate their velocity
 dispersions and also their crossing
times.  The line-of-sight velocity dispersions obtained for Sculptor and
Centaurus A are 202 \kms ~and 150 \kms ~respectively, which are typical for
groups of this size. For the crossing times estimates, 
the centers of mass were adopted to be at the mean projected position
and velocity of the most massive galaxies in each group: NGC~5128 and 
NGC~5236 for \cena A, and NGC~253 and NGC~247 for
Sculptor. NGC~45 and NGC~24, being at least at twice the distance of
the other Sculptor members, were excluded from this estimate. Although
NGC~247 is less luminous than NGC~7793, its maximum rotational velocity
 is larger. In
fact from the mass-models of  Puche \& 
 Carignan (1991) and Carignan \& Puche (1990), NGC~253 and NGC~247 together  
 provide about 90\%  of the mass of the group. 

 The indicative crossing
time (see Rood \& Dickel 1978) is then calculated by:
\begin{equation}
t_{cross} = {{2 \Bigl<r\Bigl>}\over {\pi \Bigl<|\Delta V|\Bigl>}}
\end{equation} 
where $\Bigl<r\Bigl>$ is the average projected radial distance from the center
of mass, and $\Bigl<|\Delta V|\Bigl>$ is the mean absolute radial velocity from the center 
of mass. Table~\ref{t:tcross} show the results: the crossing times are
quite 
long, $3.2 \times 10^9$ years and $4.5 \times 10^9$ years respectively, a considerable fraction
of a Hubble time, which shows that these 
two groups are probably not virialised.
  The fact that the 
groups are unvirialised is hardly surprising: for \cena A the spatial distribution
 of the objects in a long chain is easily noticeable, and for \scul the velocity
 distribution is clearly non-Gaussian. Figure~\ref{f:velhis} illustrates this point,
where cumulative velocity histograms are shown for each group. 
 The \scul histogram shows a plateau near
$|\Delta V| = 0$; ie rather few dwarfs have velocities close to the
center of mass velocity. In  
Figure~\ref{f:dvdr} we plot $|\Delta V|$ against $\Delta r$ for each object; 
here again it is verified that there is qualitatively no obvious signature of 
virialisation;  
the points cover the whole ($|\Delta V|$ --$\Delta r$) area which indicates
 that these two loose groups are still collapsing.

The situation in \scul and \cena A is however far from being unusual.
Turner \& Gott (1976) have calculated crossing times for groups identified
by Sandage, Tamman and de Vaucouleurs, and concluded that most groups are just
now entering the virialised regime. Giuricin {\etal} (1988) finds that 75\%
of the groups in Geller \& Huchra's (1983) catalogue are still in the phase
of collapse and not yet virialised. Even cluster virialisation 
 has now been put in doubt, as X-ray maps start revealing distinct
subclustering, also seen optically as more redshifts come available like in
 the Coma cluster (Colless \& Dunn 1996). Our own Local Group is certainly 
still collapsing (Gunn 1974). In fact, looking in redshift space, the
\scul group seems to form one
long appendage to the Local Group and on to the M81 group on the other side,
 as if these 3 groups were born out of the condensation of the same
 cloud, in which these 3 `clumps' have not yet quite collapsed (it is 
 referred to as the `Coma-Sculptor Cloud' in Tully's Catalog 1988). 
This makes the determination of
masses and mass-to-light ratios for these groups rather difficult. If we 
na\"{i}vely apply the virial theorem to \scul and \cena A, we obtain mass-to-light
ratios of $M/L_B = 98$ and $M/L_B = 29$ respectively (using $L_B$ from RC3 
 B values and our adopted distances of 2.5 and 3.5 Mpc).
 But if we require the groups to be bound only, their masses are then a factor  
 of two smaller than those obtained here from the virial theorem. That would  
bring the mass-to-light ratio of the whole \cena A group into a regime similar 
to mass-to-light ratios of single galaxies.
In the case of \scul however we have to
conclude that the mass-to-light ratio of the group exceeds that of individual
galaxies and suggests the presence of a large
amount of dark matter in the group.

There are more sophisticated techniques to study the dynamics of
loose groups. Peebles (1989, 1990, 1994) in a series of papers explored a new 
method of tracing nearby galaxies' orbits back in time, using a
numerical application of the action variational principle (see also 
Dunn \& LaFlamme 1993, 1995). His predicted
velocities agree well with the observed 
 velocities for the neighboring groups of galaxies within 3 Mpc  when
 they are assigned M/L of as much as $\sim $ 150 $M_{\odot} /L_{\odot}$.   
Our new dwarfs will be very valuable in testing this mass model. 

\section{Summary}

A total of 123 and 145 dwarf galaxy candidates in the two nearest groups
of galaxies Sculptor (2.5 Mpc) and Centaurus A (3.5 Mpc) were selected
visually on SRC J films, over areas of 940 and 910 square degrees
respectively. From an HI Parkes survey, with detection limits of 
$4 \times 10^6$ \Msun and $7.8 \times 10^6$ \Msun , 33 dwarf galaxies
were detected in the redshift ranges of the groups. A follow-up
H$_{\alpha }$ spectroscopic survey has confirmed 21 of these redshifts, and has
found 3 more objects not detected in HI. 

A total of 16 dwarfs are
therefore found to be members of the Sculptor group, and 20 of the
Centaurus A group.
 Most of these objects were already known in the
literature, but 5 members have newly determined redshifts and 6 other
members are identified for the first time. 

These objects are mostly 
dIrrs but with a variety of morphologies. Their global properties 
will be presented in subsequent papers.

The dwarf members have a broader distribution spatially and in velocity
than the brighter members of each group. As most nearby groups, Sculptor
and Centaurus A are not yet virialised.

\section{Acknowledgements}

It is a pleasure to thank Bruno Binggeli, Evan Skillman, and Jacqueline
van Gorkom for many interesting comments.
This survey was started at Universit\'e de Montr\'eal with the help of Serge 
Demers, Pierre Chastenay, and especially Nathalie Martimbeau who produced all  
the finding charts. 
SC thanks Tom Broadhurst for his constant criticism.
Thanks to J.English, A.Koekemoer, H.Liang, C.Lidman, R.Vaile and
especially E.Troup for their help on the Parkes observing run.  
The Parkes telescope is part of the Australia  
Telescope National Facility, which is operated in association with the Division
~of Radiophysics by CSIRO.  
This project would have been a nightmare without the help of the
NASA/IPAC Extragalactic Database (NED), which is operated by the Jet
Propulsion Laboratory, Caltech, under contract with the National
Aeronautics and Space Administration. 
Financial support for this work was provided by an Australian National
University Postgraduate Scholarship, and by Fonds FCAR Qu\'ebec.

 
\clearpage

\begin{deluxetable}{lccccccc}
\tablecaption{Main members of the Sculptor group. \label{t:sculm}}
\tablewidth{0pt}
\tablehead{
\colhead{Name  } & \colhead{ R.A. \& Dec.} & \colhead{Type} & 
\colhead{V$_{\odot}$} & \colhead{B$_T$} &
\colhead{Distance} & \colhead{M$_T$} & 
\colhead{Ref.} \\
\colhead{} & \colhead{(1950)} & \colhead{} & \colhead{(km/s)} & \colhead{} & \colhead{(Mpc)} & 
\colhead{} & \colhead{} \\   
}
\startdata
NGC 7793 & 23 55 15 -32 52 06 & SA(s)d & 230 & 9.61 & 3.38 & -18.03 & 1
\nl
NGC 55 & 00 12 24 -39 28 00 & SB(s)m  & 125 & 8.39 & 1.6 & -17.72 & 1 \nl
NGC 247 & 00 44 39 -21 02 00 & SAB(s)d & 159 & 9.60 & 2.53 & -17.41 & 1
\nl
NGC 253 & 00 45 07 -25 33 42 & SAB(s)c & 251 & 7.99 & 2.58 & -19.07 & 1
\nl
NGC 300 & 00 52 31 -37 57 24 & SA(s)d & 142 & 8.70 & 2.1 & -17.91 &  2 \nl
NGC 24 & 00 07 24 -25 14 36 & SA(s)c & 554 & 12.13 & 11.08 & -18.09 & 1
\nl
NGC 45 & 00 11 32 -23 27 35 & SA(s)dm & 468 & 11.32 & 4.35 & -16.93 & 1
\nl
\enddata
\tablecomments{The B$_T$ magnitudes are the RC3 values, corrected for
Galactic extinction. The distances are from: 
 1) Puche \& Carignan 1988 and references therein;
 2) Freedman \etal 1992.}
\end{deluxetable}

\clearpage 
 
\begin{deluxetable}{lccccccc}
\tablecaption{Main members of the Centaurus A group. \label{t:cenam}}
\tablewidth{0pt}
\tablehead{
\colhead{Name  } & \colhead{ R.A. \& Dec.} & \colhead{Type} & 
\colhead{V$_{\odot}$} & \colhead{B$_T$} &
\colhead{Distance} & \colhead{M$_T$} & 
\colhead{Ref.} \\
\colhead{} & \colhead{(1950)} & \colhead{} & \colhead{(km/s)} & \colhead{} & \colhead{(Mpc)} & 
\colhead{} & \colhead{} \\   
}
\startdata
 NGC 4945 & 13 02 31 -49 12 12 & SB(s)cd sp & 560 & 8.45 & 3.18 & -19.06 & 
 1 \nl
 NGC 5068 & 13 16 13 -20 46 36 & SAB(rs)cd & 672 & 10.37 & 5.11 & -18.17
 & 1 \nl
 NGC 5102 & 13 19 07 -36 22 12 & SA0- & 467 & 10.13 & 3.12 & -17.34 & 2
 \nl
 NGC 5128 & 13 22 32 -42 45 33 & S0 pec & 562 & 7.32 & 3.50 & -20.4 & 3
 \nl
 A1332-45 & 13 31 39 -45 17 06 & SB(s)m & 826 & 11.27 & 5.25 & -17.33 & 
 1 \nl
 NGC 5236 & 13 34 12 -29 36 48 & SAB(s)c & 516 & 8.05 & 3.70 & -19.79 & 
 1 \nl
 NGC 5253 & 13 37 05 -31 23 30 & Im pec & 404 & 10.67 & 4.09 & -17.39 & 
 4 \nl
\enddata
\tablerefs{ 1)
 de Vaucouleurs 1979; 2) McMillan \etal 1994; 3) Hui \etal 1993; 4)
 Sandage \etal 1994.}
\end{deluxetable}

\clearpage

\begin{deluxetable}{llrclc}
\tablecaption{Catalogue of dwarf candidates for the \scul group.
\label{t:sback}}
\tablewidth{0pt}
\tablehead{
\colhead{} & \colhead{} & \multicolumn{3}{c}{HI detected:} & 
 \colhead{H$\alpha$ detected:} \\
 \cline{3-5} \\
\colhead{Name  }   & \colhead{ R.A. \& Dec.} & \colhead{$V_{\odot} $} & \colhead{Int.Flux}  & \colhead{$\Delta V_{20}$} & \colhead{$V_{\odot}$} \\ 
\colhead{} & \colhead{(1950)} & \colhead{\kms }& \colhead{Jy \kms } & \colhead{\kms } & \colhead{\kms } \\  
}
\startdata
 ESO 406-G040 & 22 57 33 -37 28.1& 1248 & 4.2 & 67 & \nl
 ESO 406-G042 & 22 59 25 -37 21.2& 1377 & 7.1 & 143 & \nl
 MCG-05-54-024 & 23 11 07 -29 51.5& & & & \nl
 ESO 407-G011 & 23 12 13 -34 02.7& & & & 1869 \nl
 SC1 & 23 16 54 -31 37  & & & & \nl
 SC2  & 23 17 56 -32 10.8& 68 & 10.4 & 49 &  \nl
 ESO 347-G008 & 23 18 09 -42 00  & 1622 & 8.6 & 76 & 1606 \nl
 ESO 347-G017 & 23 24 16 -37 37.3& 702 & 10.5 & 104 & 659 \nl 
 SC3 & 23 25 51 -27 05.8& & & & \nl
 SC4 & 23 26 40 -33 35.3& & & & \nl
 ESO 470-G016 & 23 28 18 -27 56.8& & & & \nl
 MCG-05-55-020 & 23 28 21 -27 48  & & & & \nl
 SC5 & 23 28 59 -27 28.5& & & & \nl
 SC6 & 23 30 13 -28 47.2& & & & \nl
 ESO 291-G031 & 23 31 39 -46 16  & 1494 & 6.5 & 96 & \nl
 MCG-05-55-026 & 23 32 59 -26 57.5& & & & \nl
 SC7 & 23 33 48 -23 18  & & & & \nl
 ESO 292-G002 & 23 34 51 -43 53.8& & & & \nl 
 ESO 537-G001 & 23 40 23 -26 36  & & & & 3780 \nl
 UGCA 442 & 23 41 10 -32 13.8& 274 & 54.3 & 114 & 283 \nl
 SC8 & 23 43 57 -32 50.5& & & & \nl
 ESO 348-G009 & 23 46 47 -38 03  & 656 & 8.4 & 98 & 628 \nl
 SC9 & 23 48 19 -23 16.1& & & & \nl
 SC10 & 23 48 25 -29 48.5& & & & \nl
 SC11 & 23 51 01 -26 37.5& & & & 3006 \nl
 SC12 & 23 51 42 -22 57.5& & & & \nl
 SC13 & 23 53 45 -42 01.7& & & & \nl
 SC14 & 23 54 23 -31 11.2& & & & \nl
 SC15 & 23 56 03 -24 49  & & & & \nl
 SC16 & 23 56 07 -31 44.6& & & & \nl
 SC17 & 23 58 15 -26 45.4& & & & \nl
 SC18 & 23 58 22 -41 25.6& 151 & 4.6 & 46 & \nl
 ESO 409-G008 & 00 00 18 -32 08.4& & & & 7427 \nl
 SC19 & 00 00 57 -26 51.3& & & & \nl
 ESO 538-G021 & 00 03 13 -22 21.3& & & & 3148 \nl
 ESO 472-G015 & 00 04 14 -25 13.3& & & & \nl
 ESO 293-G035 & 00 04 19 -42 07  & 110 & 6.9 & 52 & \nl
 ESO 293-G040 & 00 05 00 -37 44.4& & & & \nl
 SDIG & 00 05 41 -34 51.3& 229 & 2.7 & 31 & \nl
 UGCA 003 & 00 07 45 -18 32.5& 1549 & 9.6 & 50 & 1551 \nl
 NGC 59 & 00 12 53 -21 43.2& 367 & 3.7 & 87 & 357 \nl
 ESO 410-G005 & 00 13 00 -32 27.5& & & & \nl
 ESO 194-G002 & 00 16 02 -47 56  &  & &  & 1494 \nl
 SC20 & 00 16 07 -24 17  & & & & \nl
 UGCA 005 & 00 16 17 -19 17  & & & & 3187 \nl
 SC21 & 00 19 33 -19 07.1& & & & \nl
 SC22 & 00 21 20 -24 58.7& & & & \nl
 ESO 410-G011 & 00 22 27 -27 34.2& & & & \nl
 ESO 294-G010 & 00 24 06 -42 07.8& & & & 4450 \nl
 ESO 473-G020 & 00 24 55 -25 26.5& & & & \nl
 ESO 410-G012 & 00 25 49 -28 15.3& 1545 & 5.9 & 131 & 1557 \nl
 ESO 473-G024 & 00 28 50 -23 02.6& 553 & 7.7 & 74 & \nl
 ESO 294-G020 & 00 29 45 -40 32.3& & & & 1431 \nl
 SC23 & 00 30 53 -22 08.8& & & & \nl
 ESO 410-G017 & 00 31 10 -28 04.5& 1473 & 7.0 & 126 & 1535 \nl
 SC24 & 00 34 12 -32 50.8& 79 & 11.8 & 92 & \nl
 SC25 & 00 34 33 -27 09  & & & & \nl
 AM0035-434 & 00 35 21 -43 46.5& & & & \nl
 SC26 & 00 37 44 -18 05.7& & & & \nl
 ESO 540-G012 & 00 37 49 -20 58  & & & & 3933 \nl
 SC27 & 00 38 32 -26 32.5& & & & 2694 \nl
 DDO 226 & 00 40 36 -22 31.4& 372 & 7.5 & 59 & \nl
 SC28 & 00 41 12 -17 47.8& & & & \nl
 SC29 & 00 42 43 -25 32.4& & & & \nl
 ESO 411-G013 & 00 44 42 -31 51  & & & & \nl
 ESO 411-G018 & 00 45 35 -32 15  & 1728 & 5.2 & 161 & 1712 \nl
 SC30 & 00 45 56 -25 44.5& & & & \nl
 ESO 411-G019 & 00 46 41 -29 23.6& & & & 1817 \nl
 ESO 540-G030 & 00 46 53 -18 20.8& & & & \nl
 0047-21 & 00 47 18 -21 56.8& & & & \nl
 DDO 006 & 00 47 20 -21 17.5& 304 & 4.5 & 58 & \nl
 ESO 540-G032 & 00 47 56 -20 10.8& & & & \nl
 ESO 411-G027 & 00 50 26 -27 35.8& & & & 1829 \nl
 SC31 & 00 51 11 -25 44.9& & & & \nl
 ESO 540-G033 & 00 51 22 -19 49  & & & & 6398 \nl
 ESO 351-G029 & 00 57 46 -35 48.6& & & & \nl
 ESO 541-IG012 & 00 59 51 -19 56  & & & & 16954  \nl
 AM0106-382 & 01 06 06 -38 28.4&  &  &  & 645 \nl
 ESO 195-G032 & 01 07 19 -48 03.5& & & & 7685 \nl
 ESO 243-G050 & 01 08 34 -42 38.5& 1477 & 2.2 & 72 & 1490 \nl
 SC32 & 01 10 27 -38 31.1& & & & 6607 \nl
 SC33 & 01 15 20 -21 56.2& & & & \nl
 SC34 & 01 19 44 -27 39.7& & & & \nl
 SC35 & 01 20 11 -26 16  & & & & \nl
 ESO 476-G010 & 01 24 23 -25 34.4& 1607 & 4.2 & 134 & \nl
 SC36 & 01 24 31 -15 48.5& & & & \nl
 SC37 & 01 26 32 -41 31.8& & & & \nl
 ESO 542-G022 & 01 28 50 -17 57.5& & & & 5442 \nl
 SC38 & 01 30 33 -39 46.5& & & & \nl
 SC39 & 01 31 27 -29 57  & & & & \nl
 ESO 476-G023 & 01 31 29 -25 28  & & & & \nl
 SC40 & 01 31 31 -19 19.7& & & & \nl
 ESO 476-G023 & 01 31 31 -25 27.9& & & & \nl
 ESO 353-G017 & 01 31 50 -34 41.9& & & & 3820 \nl
 NGC 625 & 01 32 56 -41 41.4& 406 & 32.5 & 106 & 415 \nl
 SC41 & 01 35 20 -34 48.7& & & & \nl
 SC42 & 01 37 11 -47 33.1& 162 & 8 & 64 & \nl
 MCG-03-05-014 & 01 39 09 -16 23.8& 1632 & 5.9 & 143 & \nl
 SC43 & 01 41 51 -27 55.5& & & & \nl
 ESO 245-G005 & 01 42 58 -43 50.5& 399 & 87.3 & 94 & 389 \nl
 SC44 & 01 48 00 -17 02.7& & & & \nl
 Phoenix & 01 49 02 -44 41.3& 56 & 2.4 & 29 & \nl
 ESO 477-G012 & 01 51 28 -23 21.1& & & & 1463 \nl
 ESO 477-G017 & 01 54 36 -25 46  & & & & \nl
\enddata
\end{deluxetable}

\clearpage

\begin{deluxetable}{llrclc}
\tablecaption{Catalogue of dwarf candidates for the \cena A group.
\label{t:cback}}
\tablewidth{0pt}
\tablehead{
\colhead{} & \colhead{} & \multicolumn{3}{c}{HI detected:} & 
 \colhead{H$\alpha$ detected:} \\
 \cline{3-5} \\
\colhead{Name  }   & \colhead{ R.A. \& Dec.} & \colhead{$V_{\odot} $} & \colhead{Int.Flux}  & \colhead{$\Delta V_{20}$} & \colhead{$V_{\odot}$} \\ 
\colhead{} & \colhead{(1950)} & \colhead{\kms }& \colhead{Jy \kms } & \colhead{\kms } & \colhead{\kms } \\  
}
\startdata
 ESO 321-G018 & 12 13 10 -37 50  & & & & 3163 \nl
 ESO 322-G015 & 12 24 57 -37 50  & & & & \nl
 ESO 574-G001 & 12 25 55 -21 57.5& & & & 6763 \nl
 ESO 218-G012 & 12 37 10 -51 57.5& & & & 1809 \nl
 ESO 268-G033 & 12 39 45 -47 17  & & & & 5437 \nl
 ESO 172-G006 & 12 40 21 -52 30.5& & & & 1910 \nl
 ESO 381-G020 & 12 43 16 -33 34  & 596 & 31.9 & 103 & 599 \nl
 ESO 443-G001 & 12 51 09 -27 36  & & & & 3187 \nl
 ESO 443-G006 & 12 51 55 -31 36  & & & & 3563 \nl
 CEN1 & 12 53 50 -42 02  & & & & \nl
 ESO 381-G031 & 12 55 30 -32 56  & & & & 2467 \nl
 ESO 323-G054 & 12 56 35 -40 42  & & & & 2944 \nl
 ESO 269-G024 & 12 56 40 -42 53  & & & & 3584 \nl
 ESO 219-G017 & 12 56 55 -47 59  & & & & 3532 \nl
 ESO 323-G056 & 12 57 20 -38 08  & & & & 4550 \nl
 CEN2 & 12 58 05 -47 26.5& & & & 2115 \nl
 CEN3 & 12 58 12 -47 27.3& & & & \nl
 ESO 507-G065 & 12 58 21 -25 49  & & & & 2607 \nl
 CEN4 & 12 58 50 -25 55  & & & & \nl
 SGC1259.6-1659 & 12 59 33 -16 58  & 732 & 5.2 & 63 & 720 \nl
 DDO 161 & 13 00 37 -17 09  & 747 & 110.1 & 136 & 750 \nl
 CEN5 & 13 02 02 -49 08  & 122 & 5.5 & 53 & \nl
 CEN6 & 13 02 12 -39 48  & 619 & 4.4 & 43 & 602 \nl
 ESO 219-G027 & 13 03 39 -49 35  & 1281 & 11.7 & 112 \nl
 ESO 219-G028 & 13 03 41 -49 25  & 1438 & 9.9 & 143 & \nl
 ESO 508-G004 & 13 04 10 -22 34  & & & & 2896 \nl
 ESO 269-G053 & 13 05 57 -42 39  & & & & 2002\nl
 ESO 443-G075 & 13 06 25 -28 22  & 1425 & 25 & 172 \nl
 ESO 219-IG032 & 13 06 30 -47 42.7& & & & 3489 \nl
 UKS 1307-429 & 13 07 10 -42 56  & & & & 2126 \nl
 ESO 443-G079 & 13 07 38 -27 42  & & & & 2138 \nl
 CEN7 & 13 08 24 -38 38 & & & & \nl
 ESO 443-G080 & 13 08 21 -27 44  & & & & 2137 \nl
 ESO 443-G083 & 13 10 10 -32 25  & & & & 2389 \nl
 ESO 323-G097 & 13 11 10 -39 00  & & & & 5010 \nl
 ESO 508-G030 & 13 12 12 -22 52  & 1502 & 19.3 & 151 & \nl
 ESO 443-G085 & 13 12 25 -32 00  & & & & 2384 \nl
 ESO 444-G002 & 13 14 00 -27 37  & 1636 & 10.2 & 122 & \nl
 ESO 382-G030 & 13 14 10 -37 24  & & & & \nl
 ESO 444-G006 & 13 15 43 -31 21  & & & & 3653 \nl
 ESO 382-G040 & 13 16 35 -33 00.5& 1682 & 14.5 & 144 & \nl
 ESO 269-G087 & 13 17 20 -47 25.5& & & & 3000 \nl
 ESO 382-G045 & 13 17 25 -35 46  & 1460 & 33.6 & 155 & \nl
 AM 1317-425 & 13 17 42 -42 50  & & & & 3323  \nl
 CEN8 & 13 20 04 -33 17   & & & & \nl
 AM1321-304 & 13 21 49 -30 42   & & & & \nl
 ESO 382-G061 & 13 22 41 -37 07  & 1437 & 7.3 & 105 & \nl
 CEN9 & 13 22 48 -29 49  & & & & 4413 \nl
 ESO 576-G059 & 13 23 52 -21 57.5& 1431 & 10.2 & 118 & \nl
 ESO 220-G011 & 13 24 00 -48 28  & & & & 2925 \nl
 ESO 324-G023 & 13 24 36 -37 55  & 1439 & 71.9 & 218 & \nl
 ESO 324-G024 & 13 24 40 -41 13  & 526 & 52.1 & 113 & 524 \nl
 CEN10 & 13 26 06 -30 13.5& & & & \nl
 ESO 444-G059 & 13 27 42 -31 57  & & & & \nl
 ESO 383-G017 & 13 30 25 -34 12  & & & & 3471 \nl
 NGC 5206 & 13 30 41 -47 53.7& & & & 571 \nl
 ESO 509-G059 & 13 31 05 -24 30  & & & & 4664 \nl
 Fourc-Figu & 13 31 45 -45 16  & 831 & 201.6 & 157 & \nl
 CEN11 & 13 31 48 -28 58.5& & & & 1347 \nl
 UGCA 365 & 13 33 42 -28 59  & 582 & 4.6 & 61 & \nl
 ESO 444-G084 & 13 34 30 -27 47  & 591 & 19.6 & 82 & 578 \nl
 NGC 5237 & 13 34 40 -42 35.5& 369 & 7.6 & 89 & 371 \nl
 ESO 324-G044 & 13 35 15 -39 35  & & & & 2532 \nl
 IC 4316 & 13 37 28 -28 38  & 589 & 7.8 & 120 & 589 \nl
 NGC 5264 & 13 38 47 -29 39.7& 487 & 13.7 & 84 & 476 \nl
 ESO 220-G032 & 13 38 58 -48 04  & 1364 & 7.7 & 114 & \nl
 ESO 383-G062 & 13 39 05 -35 26  & & & & \nl
 ESO 383-G064 & 13 39 15 -36 06  & & & & \nl
 ESO 509-G096 & 13 40 07 -24 52  & & & & 5467 \nl
 ESO 445-G023 & 13 41 52 -29 49  & & & & 4620 \nl
 ESO 325-G011 & 13 42 00 -41 36  & 550 & 25.4 & 77 & 554 \nl
 ESO 383-G087 & 13 46 20 -35 50  & 333 & 27.4 & 68 & 342 \nl
 ESO 383-G091 & 13 47 36 -37 03  & 1077 & 9.4 & 155 & \nl
 ESO 384-G002 & 13 48 28 -33 34  & 1391 & 65.5 & 146 & \nl
 ESO 445-G061 & 13 49 15 -31 35  & & & & \nl
 AM1357-504 & 13 57 30 -50 48.5& & & & -48\tablenotemark{a} \nl
 CEN12 & 14 00 15 -50 31.5& & & & \nl
 NGC 5408 & 14 00 17 -41 09  & 506 & 65.5 & 123 & 497 \nl
 ESO 446-G020 & 14 06 38 -30 02  & & & & 2629 \nl
 ESO 579-G003 & 14 10 48 -17 45  & 1445 & 20.5 & 170 & \nl
 ESO 271-G025 & 14 11 50 -43 43  & 1786 & 25.2 & 116 & \nl
 ESO 222-G001 & 14 15 30 -47 30  & 1284 & 22.1 & 170 & \nl
 ESO 272-G004 & 14 16 00 -45 05  & 1653 & 20.5 & 171 & \nl
 ESO 579-G019 & 14 16 43 -21 31  & & & & \nl
 ESO 446-G053 & 14 18 22 -29 02  & 1391 & 23.2 & 169 & \nl
 ESO 579-G021 & 14 19 15 -18 44  & & & & \nl
 ESO 385-G014 & 14 19 20 -35 48  & & & & 3659 \nl
 ESO 222-G004 & 14 20 24 -49 26  & 1555 & 26.9 & 115 & \nl
 ESO 385-G019 & 14 20 45 -35 18  & & & & 3493 \nl
 SGC 142448-4604.8 & 14 24 48 -46 04.8& 397 & 19.5 & 68 & \nl
 ESO 511-G050 & 14 28 40 -25 10  & & & & 2555 \nl
 ESO 222-G010 & 14 31 42 -49 12.5& 632 & 9.1 & 83 & 603 \nl
 CEN13 & 14 35 15 -46 40  & & & & \nl
 IC 4472 & 14 36 45 -44 06  & & & & 2869 \nl
 ESO 272-G025 & 14 40 10 -44 29  & & & & 624 \nl
 ESO 222-G015 & 14 41 05 -49 11  & & & & 2254 \nl
 ESO 386-G022 & 14 43 52 -35 18  & & & & 6120 \nl
 IC 4501 & 14 44 34 -22 12  & & & & 3319 \nl
 CEN14 & 14 45 45 -28 07  & & & & \nl
 SGC145501-4730 & 14 55 05 -47 28  & 1056 & 155 & 205 & \nl
 ESO 223-G009 & 14 57 35 -48 04  & 593 & 96.2 & 103 & 593 \nl
 NGC 5810 & 14 59 54 -17 40  & & & & 3297 \nl
 ESO 274-G001 & 15 10 45 -46 36.5& 528 & 117 & 177 & 507 \nl
\enddata
\tablenotetext{a}{Planetary Nebula}
\end{deluxetable}

\clearpage

\begin{deluxetable}{llr}
\tablecaption{Confirmed dwarf galaxies in the \scul group.
\label{t:sfin}}
\tablewidth{0pt}
\tablehead{
\colhead{Name  } &  \colhead{ R.A. \& Dec.} & \colhead{M$_{HI}$} \\ 
\colhead{} & \colhead{(1950)} & \colhead{($10^6 M_\odot$)} \\  
}
\startdata
SC2 & 23 17 56 -32 10 49 & 15.3\nl
ESO 347-G017 & 23 24 17 -37 37 18 & 15.4\nl
UGCA 442 & 23 41 10 -32 13 58 & 82.6\nl
ESO 348-G009 & 23 46 47 -38 02 55 & 12.4\nl
SC18 & 23 58 22 -41 25 40 & 6.8\nl
ESO 293-G035 & 00 04 19 -42 07 01 & 10.2\nl
SDIG & 00 05 41 -34 51 24 & 4.0\nl
NGC 59 & 00 12 53 -21 43 12 & 5.4\nl
ESO 473-G024 & 00 28 50 -23 02 37 & 11.3\nl
SC24 & 00 34 12 -32 50 52 & 17.4\nl
DDO 226 & 00 40 36 -22 31 20 & 11.0\nl
DDO 006 & 00 47 20 -21 17 27 & 6.6\nl
AM0106-382 & 01 06 06 -38 28 21 & $<$2.8\nl
NGC 625 & 01 32 56 -41 41 24 & 47.8\nl
SC42 & 01 37 11 -47 33 10 & 11.8\nl
ESO 245-G005 & 01 42 58 -43 50 33 & 128.4\nl
\enddata
\end{deluxetable}

\clearpage

\begin{deluxetable}{llr}
\tablecaption{Confirmed dwarf galaxies in the \cena A group.
\label{t:cfin}}
\tablewidth{0pt}
\tablehead{
\colhead{Name  } &  \colhead{ R.A. \& Dec.} & \colhead{M$_{HI}$} \\ 
\colhead{} & \colhead{(1950)} & \colhead{($10^6 M_\odot$)} \\  
}
\startdata
ESO 381-G020 & 12 43 18 -33 33 54 & 92.0\nl
SGC1259.6-1659 & 12 59 35 -16 58 13 & 15.0\nl
DDO 161 & 13 00 38 -17 09 14 & 317.5\nl
CEN5 & 13 02 02 -49 08 00 & 15.9\nl
CEN6 & 13 02 12 -39 48 00 & 12.7\nl
ESO 324-G024 & 13 24 40 -41 13 18 & 150.2\nl
NGC 5206 & 13 30 41 -47 53 42 & $<$5.5\nl
UGCA 365 & 13 33 42 -28 58 54 & 13.3\nl
ESO 444-G084 & 13 34 32 -27 47 30 & 56.5\nl
NGC 5237 & 13 34 40 -42 35 36 & 21.9\nl
IC 4316 & 13 37 29 -28 38 30 & 22.5\nl
NGC 5264 & 13 38 47 -29 39 42 & 39.5\nl
ESO 325-G011 & 13 42 01 -41 36 30 & 73.2\nl
ESO 383-G087 & 13 46 23 -35 48 48 & 79.0\nl
NGC 5408 & 14 00 18 -41 08 12 & 188.9\nl
SGC142448-4604.8 & 14 24 48 -46 04 48 & 56.2\nl
ESO 222-G010 & 14 31 41 -49 12 12 & 26.2\nl
ESO 272-G025 & 14 40 09 -44 29 36 & $<$5.5\nl
ESO 223-G009 & 14 57 35 -48 04 00 & 277.4\nl
ESO 274-G001 & 15 10 45 -46 36 30 & 337.4\nl
\enddata
\end{deluxetable}

\clearpage

\begin{deluxetable}{lccr}
\tablecaption{Crossing times for the \scul and \cena A groups
\label{t:tcross}}
\tablewidth{0pt}
\tablehead{
\colhead{ } & \colhead{$\Bigl<r\Bigl>$} 
 & \colhead{$\Bigl<|\Delta V|\Bigl>$} & \colhead{$t_{cross}$} \\ 
\colhead{} & \colhead{(Mpc)} & \colhead{(\kms)} & \colhead{(years)} \\  
}
\startdata
 {\it Sculptor} & 0.66 & 127 & 3.24 $\times$ 10$^9$ \nl
 {\it Centaurus A} & 0.72 & 100 & 4.48 $\times$ 10$^9$ \nl
\enddata
\end{deluxetable}

\clearpage


{}

\clearpage

\figcaption{ HI Parkes spectra of detections in the \scul 
 region \label{f:sallsp}}
\figcaption{ HI Parkes spectra of detections in the \cena A
 region \label{f:callsp}}

\figcaption{ Atlas of dwarf galaxies in the \scul and \cena A groups,
from enlarged SRC J plates (always North at the top, East to the left). The 
scale is indicated on the first Plate for the Sculptor galaxies and on the last
for the Centaurus A ones. \label{f:atlas}}

\figcaption{ Distribution in R.A.-Dec of the \scul ({\it top}) and 
\cena A ({\it bottom}) groups members (the small filled circles indicate
the dwarf members). \label{f:dis}}

\figcaption{ Declination versus $V_{LG}$ (velocity relative to the
Local Group, in \kms ) for members of the groups and for background detections.
Triangles indicate main members, filled squares dwarf members and open
squares background objects. \label{f:vlg}}

\figcaption{ Cumulative velocity histograms, where 
 $\Delta V$ is 
the velocity of a dwarf member relative to that of the center of mass of
the group. \label{f:velhis}}

\figcaption{ For each dwarf galaxy is plotted its relative 
velocity versus its projected distance 
from the
center of mass of the Sculptor group (left) and the Centaurus A group
(right).\label{f:dvdr}}

\clearpage

\begin{figure}
\plotone{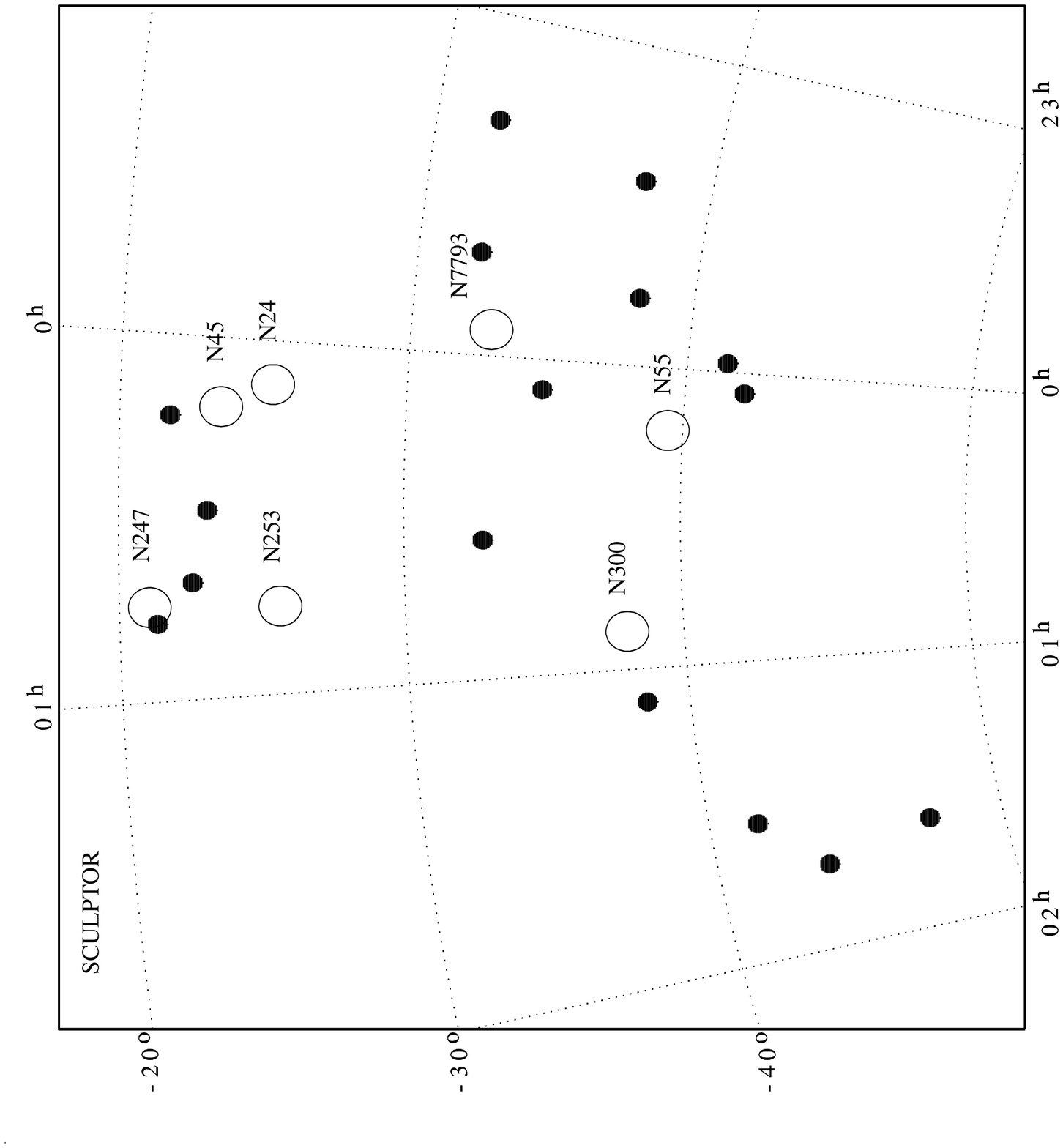}
\end{figure}

\clearpage

\begin{figure}
\plotone{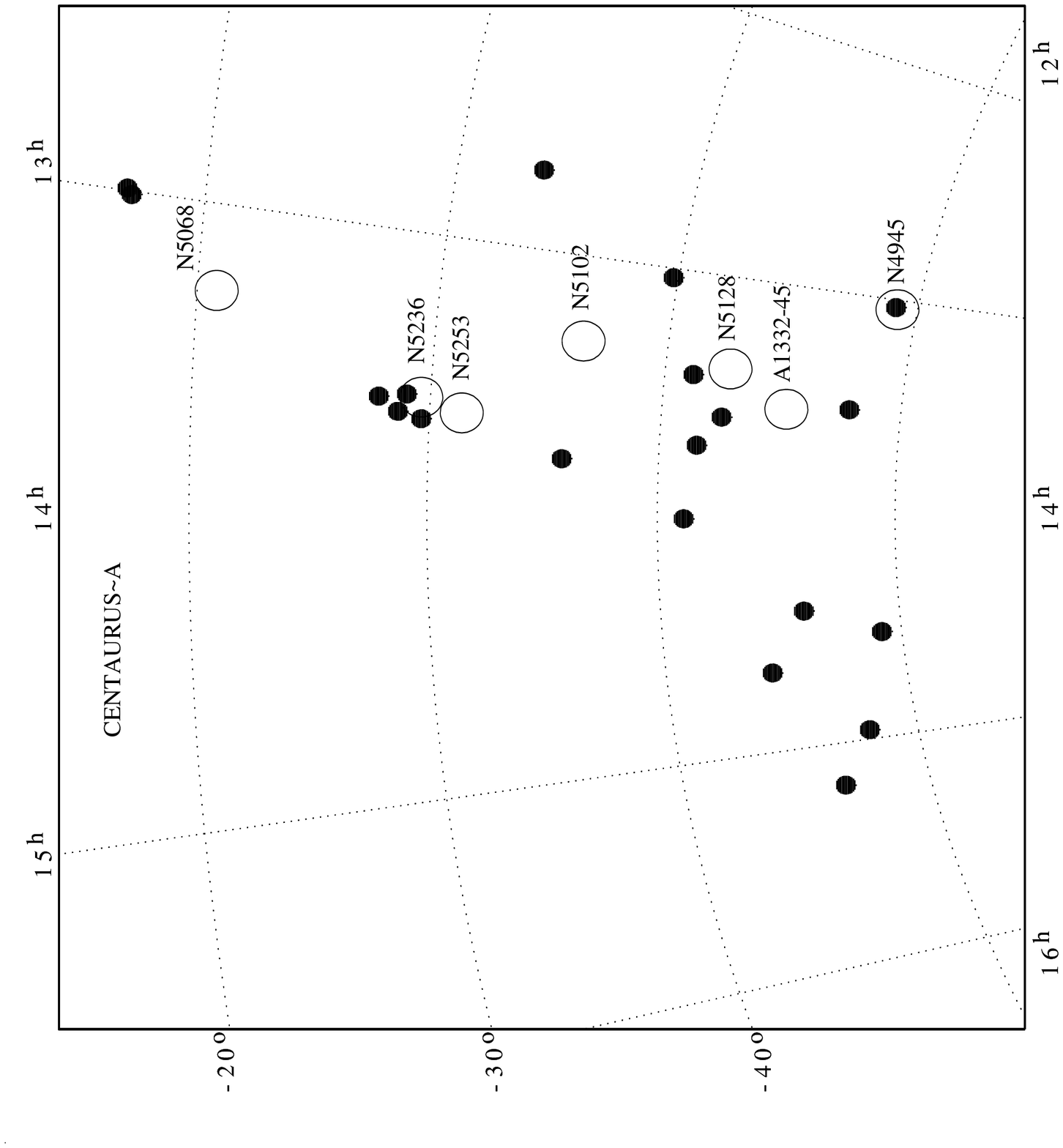}
\end{figure}

\clearpage
\begin{figure}
\plotone{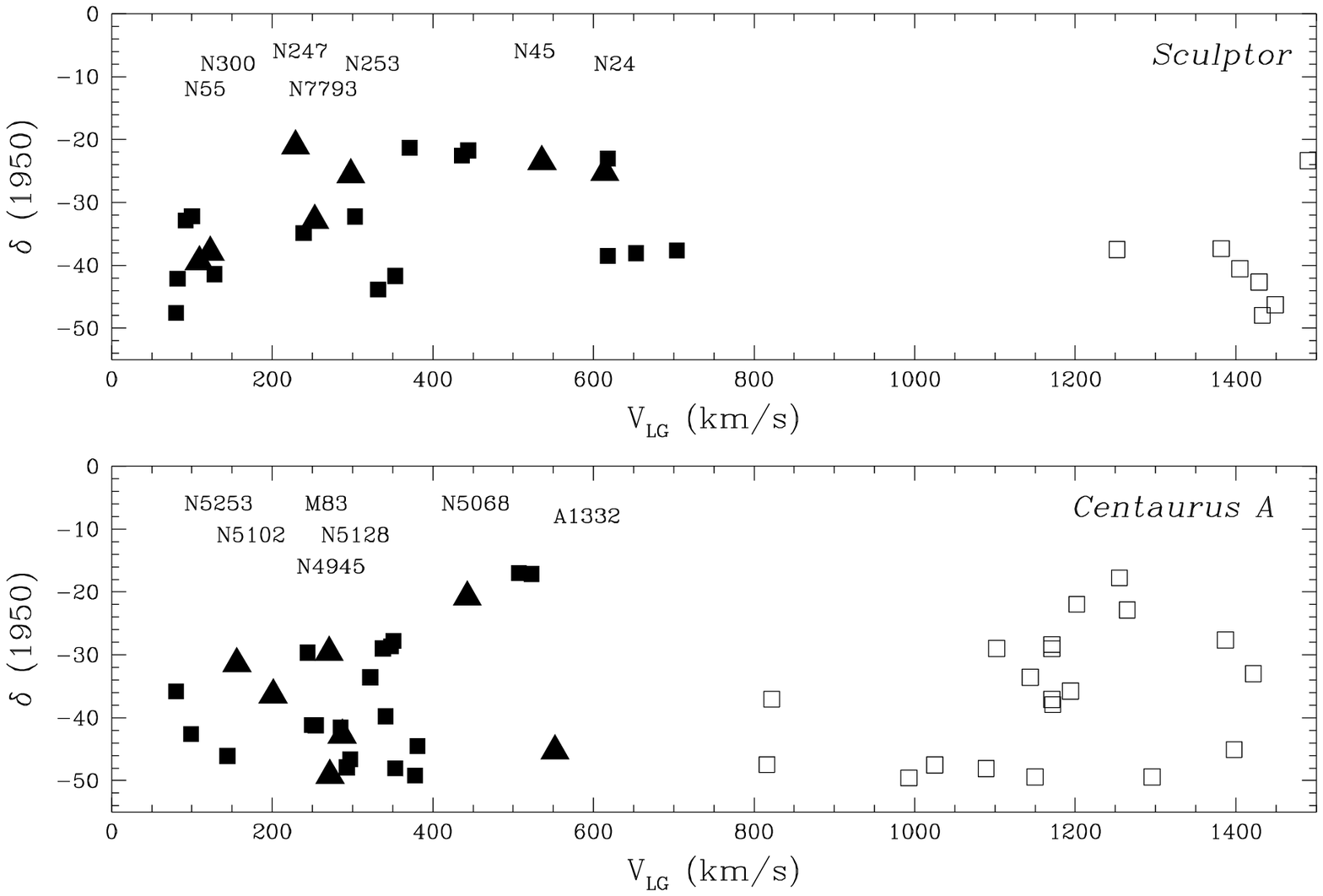}
\end{figure}

\clearpage

\begin{figure}
\plottwo{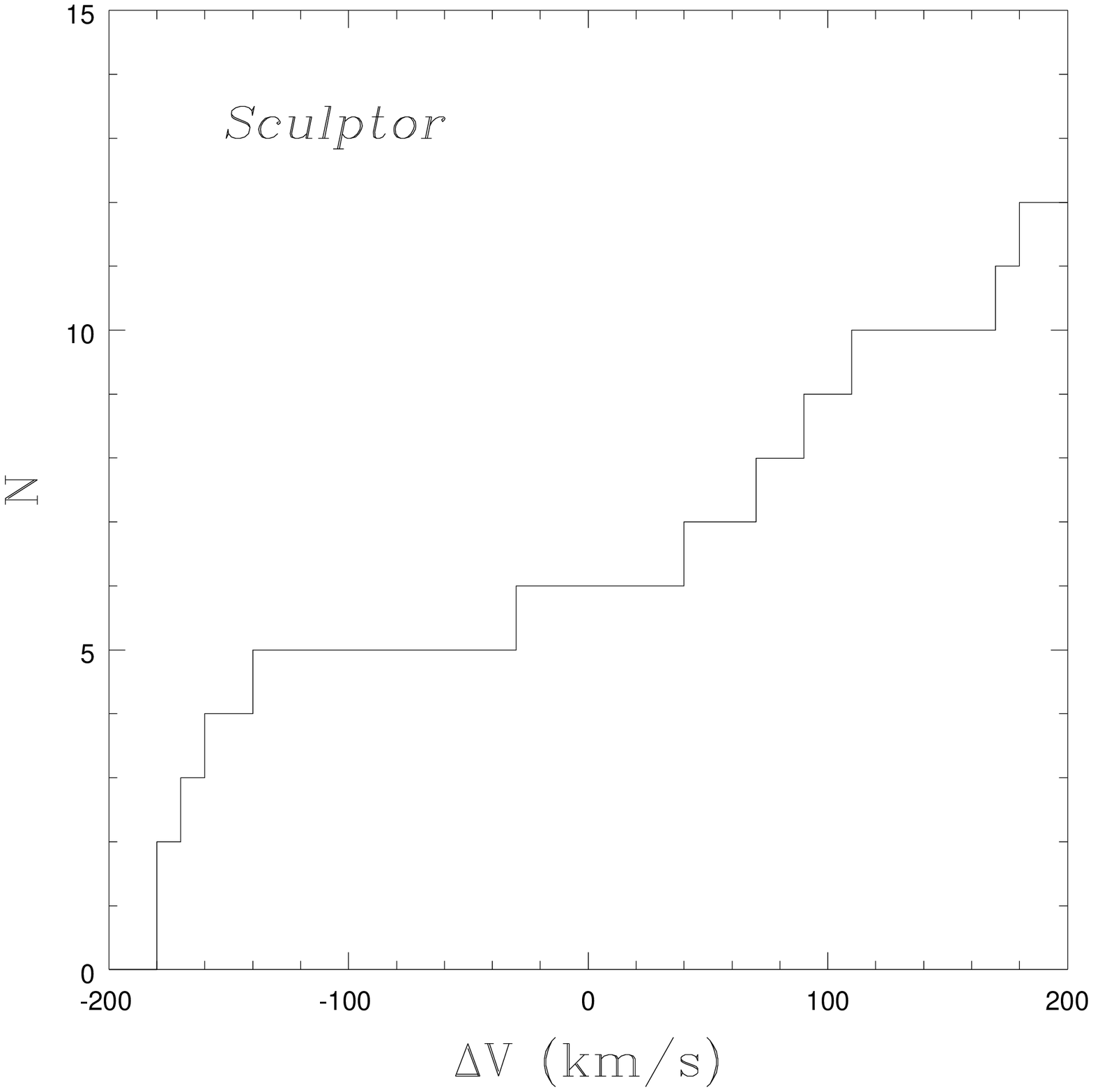}{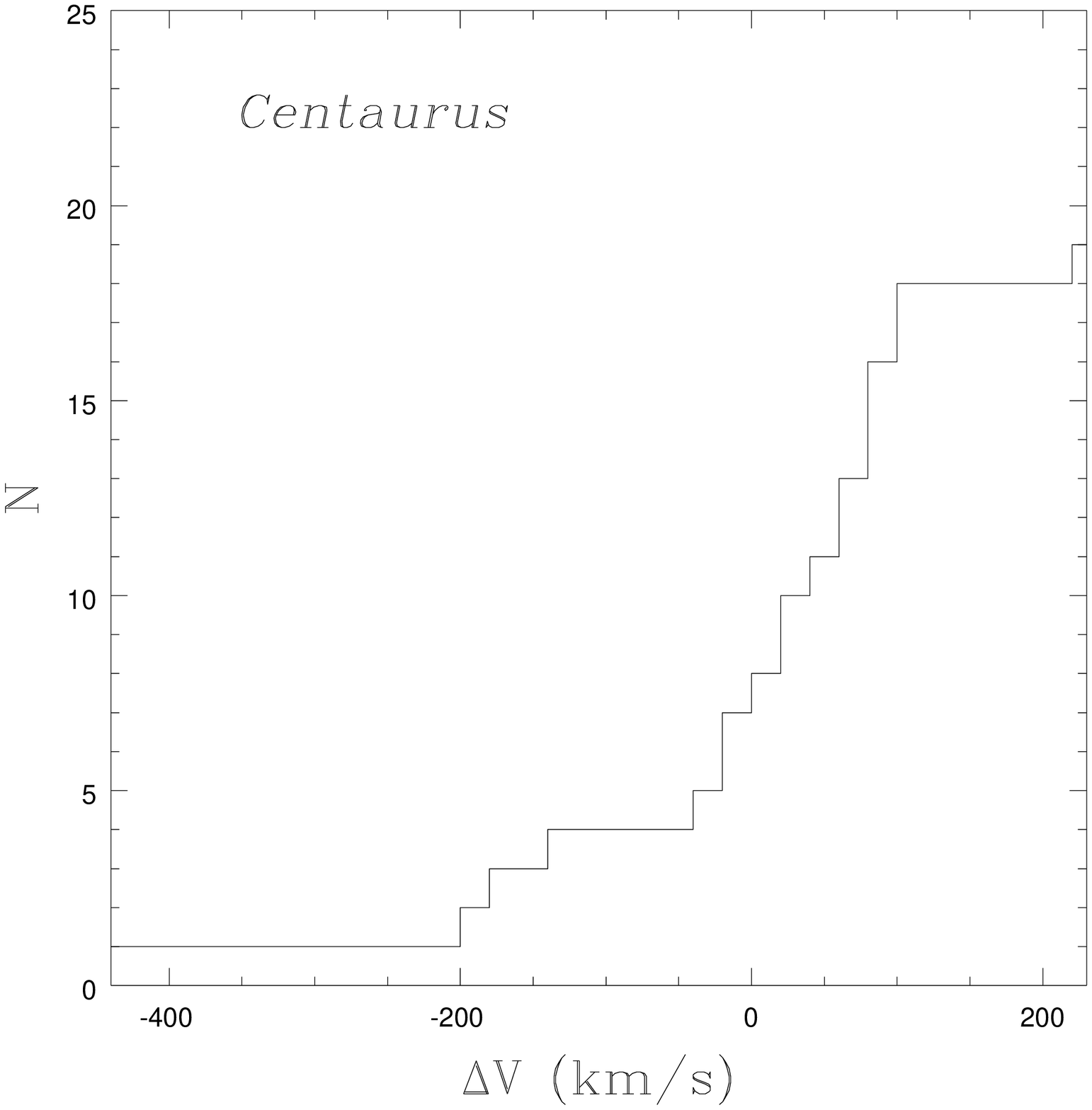}
\end{figure}

\begin{figure}
\plottwo{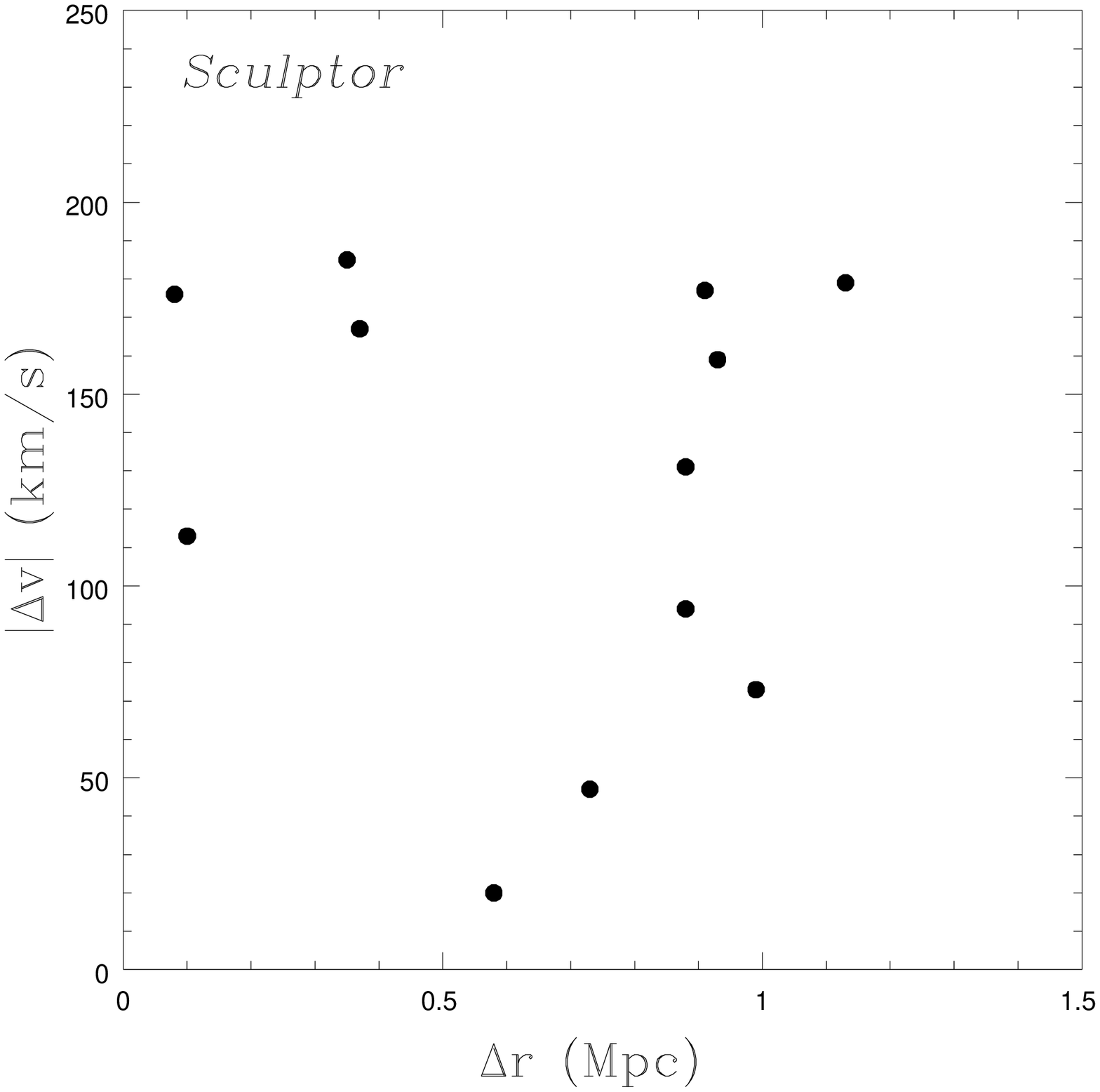}{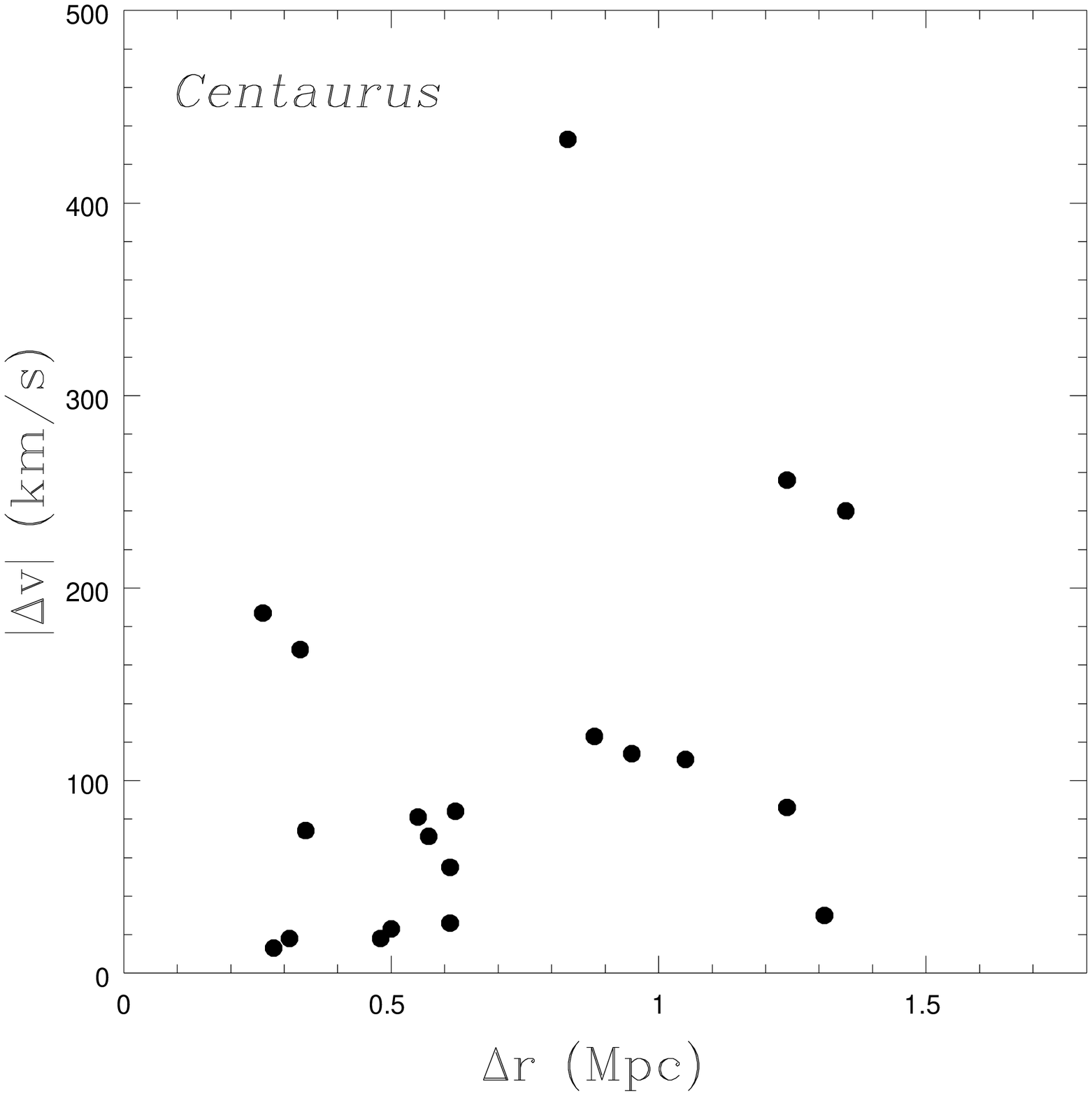}
\end{figure}

\end{document}